\documentclass{article}
\usepackage{graphicx}

\begin{document}

\title{\bf Quantum cosmology and the accelerated Universe\footnote{\uppercase{T}his talk is based on a work 
in collaboration with \uppercase{N}. \uppercase{P}into-\uppercase{N}eto, see \uppercase{R}ef. [1]}}

\author{E. Sergio Santini\footnote{\tt{santini@cbpf.br}} \\ Centro Brasileiro de Pesquisas F\'{\i}sicas \\ 
Rua Dr. Xavier Sigaud 150, CEP:22290-180, Rio de Janeiro, Brasil  \and 
Comiss\~ao Nacional de Energia Nuclear \\Rua General Severiano 90, CEP 22294-900, Rio de Janeiro, Brasil \and  
Talk given at X Marcell Grossmann Meeting, Rio de Janeiro 2003}

\date{}
\maketitle

\begin{figure}[ht]
\includegraphics{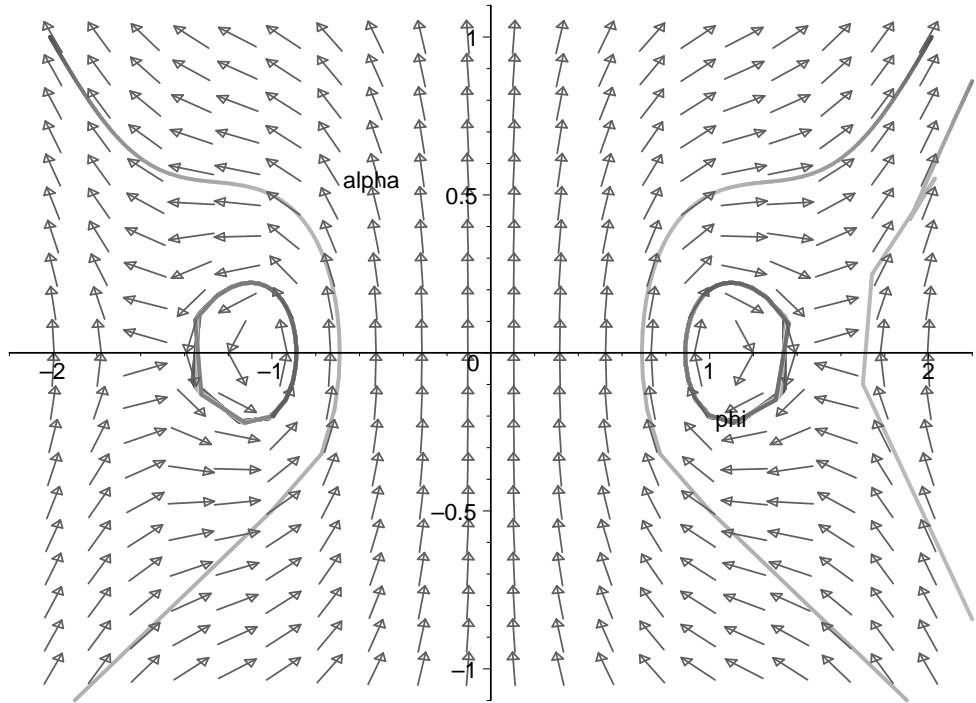}
\caption{Field plot of the system of planar equations (the guidance equations),
giving the quantum trajectories. For numerical simplicity we choose 
the values $-d=\sigma=1$.}
\label{configu}
\end{figure}

\begin{figure}[ht]
\includegraphics{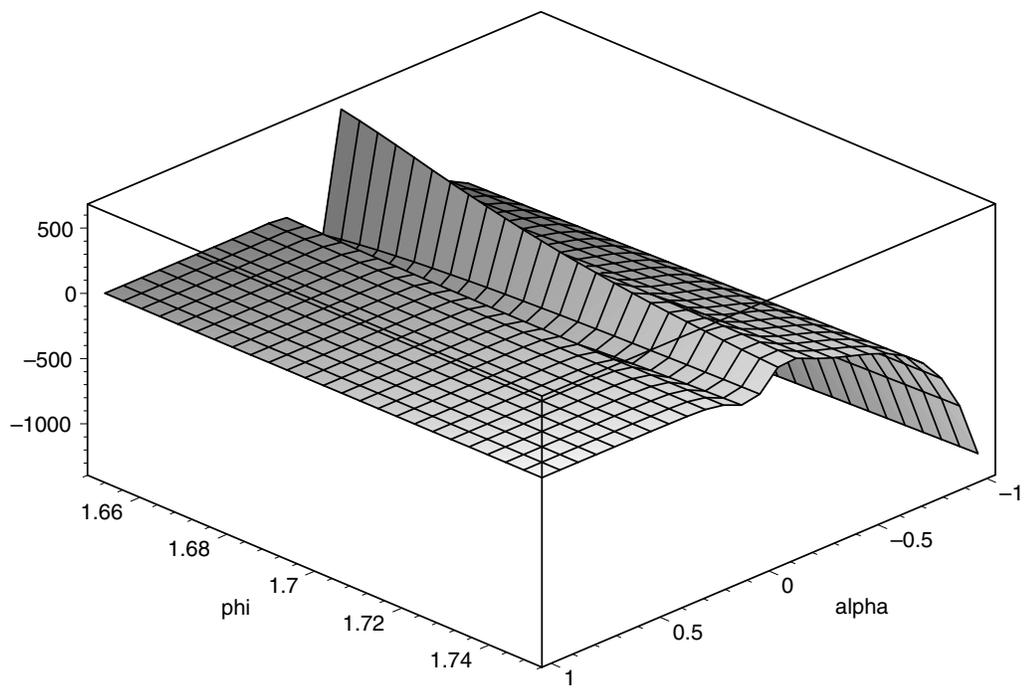}   
\caption{Acceleration $\ddot{a}/a$ as a function of $\phi$ and $\alpha$. For numerical simplicity we choose 
the values $-d=\sigma=1$.}
\label{ace}
\end{figure}

\begin{figure}[ht]
\includegraphics{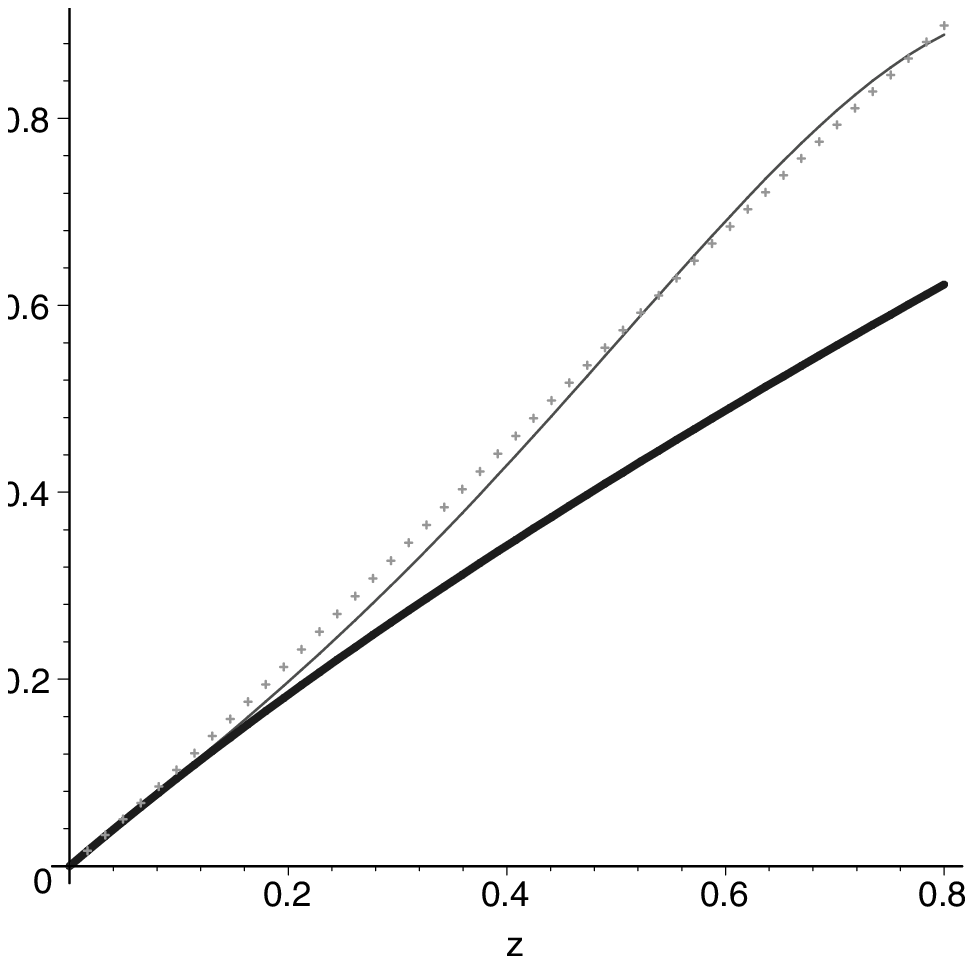}
\caption{The luminosity distance as a function of  redshift. The thin line
curve corresponds to the quantum model, the dotted curve is for the
classical model with a cosmological constant, and the thick line curve
is for the classical model without cosmological constant.}
\label{compare2}
\end{figure}

In order to explain the present accelerated expansion of the Universe, suggested by the measurements of high 
redshifts supernovae\cite{SN1}~\cite{SN}, we consider that presently, at cosmological scales, classical General 
Relativity (GR) is not valid. Instead of changing the right hand side of Einstein's equations by introducing some new 
 negative pressure fluid, we modify its left hand side according to physical considerations.
 We study a quantum universe and, keeping in mind that a macroscopic universe behaves classicaly or quantically 
 depending on its initial quantum state, we investigate if is it possible that quantum cosmological effects at large scales 
  can mimic a negative pressure fluid yielding  a positive acceleration. The aim of this talk is to show with a simple model 
  that it is indeed possible for some suitable initial quantum states of the universe.
  We consider a quantum minisuperspace model containing a free massless scalar field $\phi$  minimally coupled to gravity in a Friedmann geometry. We interpret this model in the framework of the Bohm-de Broglie (BdB) interpretation of quantum theories 
 \cite{bohm1}  \cite{hol}, in order to extract predictions from the wave functional of the Universe\footnote{Please 
 see Ref. \cite{fab2} for a previous work on BdB interpretation of quantum minisuperspace models and Ref.\cite{must}\cite{cons}\cite{tese} 
 for the full quantum superspace. }. After quantize the model according to Dirac's procedure, we obtain the Wheeler-DeWitt (WDW) equation.
We only concentrate in the flat model and
we consider the following solution of WDW equation $\Psi =  \mid \sigma \mid \sqrt{\pi}\{\exp{
[ -\frac{(\alpha+\phi)^2\sigma^2}{4}]} 
\exp(id(\alpha+\phi)) 
+\exp{[ -\frac{(\alpha-\phi)^2\sigma^2}{4}]}
\exp(id(\alpha-\phi))\}$,
where $\alpha \equiv \ln(a)$ being  $a$ the (dimensionless) scale factor and $\sigma$ and $d$ are parameters of the quantum model\cite{acel}.
The quantum trajectories, can be obtained by integrating the Bohm's guidance equations, given by: $\dot{\alpha}=\frac{\phi \sigma^2 \sin(2d\phi)-2d \cos(2d\phi)-2d \cosh(\sigma^2\alpha\phi)}
{e^{3\alpha}2[\cos(2d\phi)+\cosh(\sigma^2\alpha\phi)]}$ and $\dot{\phi}=-\frac{\alpha\sigma^2 \sin(2d\phi)+2d \sinh(\sigma^2\alpha\phi)}
{e^{3\alpha}2[\cos(2d\phi)+\cosh(\sigma^2\alpha\phi)]}$. They are shown in Fig. \ref{configu}.
We can see two different possibilities.
Oscillating universes without
singularities around the centers points and non-oscillating universes.
A non oscillating universe arises classically from a singularity, experiences
quantum effects
in the middle of its expansion, and recover  its classical behaviour for large
values of $\alpha$. The quantum effects appearing in the middle of the expansion  can deviate it from its classical decelerated
expansion to an accelerated one. We can  see that this is indeed the case for this model, by plotting the acceleration $\frac{\ddot{a}}{a}$ as a 
function of $\alpha$ and $\phi$ : $\frac{\ddot{a}}{a}=E(\alpha,\phi)$\footnote{The expression for $E(\alpha,\phi)$ can be found in Ref.\cite{acel}.} 
(Fig. \ref{ace}).
 We can see regions on the plane $\alpha$-$\phi$ 
where the acceleration is positive, negative or zero.
One can see the classical behaviour $\ddot{a}/a
 \propto -1/a^6
$ for $ a \rightarrow 0$ ($\alpha \rightarrow -\infty$) and
$a \rightarrow \infty$ ($\alpha \rightarrow \infty$), but near the region
$a=1$ ($\alpha=0$), a clear departure from classical
behaviour is observed, and positive values of $\ddot{a}/a$ are obtained.
We can compare the quantum cosmological model with the original classical free scalar 
field model, classically equivalent to stiff matter, with flat spatial section, suplemented 
with a cosmological constant as an alternative source for accelerated expansion. 
It is possible to show that for the quantum model we have  $-0.21<q_0<-0.17$ for $2d\phi_0=2n\pi+x$
with $x \in (2.145,2.15)$, provided a very large value for $\sigma$\cite{acel}. Note that,
for the classical model
with $\Omega_{\Lambda}=0.73$, $q_0=-0.19$.
The supernovae measurements relate the luminosity distance $d_L$ with $z$.
Hence, it would be instructive to compare the quantum cosmological
luminosity distance $d_L^q(z)=(1+z)\int_0^z\frac{{\rm d}y }{\dot{\alpha}(y)}$, with the  classical one.
In Fig.\ref{compare2}, we show a plot of $H_0 d_L(z)$ with $\Omega_{\Lambda}=0.73$ ,
$H_0 d_L(z)$ with $\Omega_{\Lambda}=0$, and $H_0 d_L^q(z)$.
  Note that for small values of  $z$ they are close but,
for intermediary values of $z$, the quantum $d_L^q(z)$ remain close to the
cosmological constant $d_L(z)$ while both separates of the pure stiff matter
$d_L(z)$. 
Hence, quantum  cosmological
effects may mimic a cosmological constant in some region but not everywhere.
In this way, it may be possible to explain the positive acceleration
suggested by the recent measurements of high redshift supernovae
without postulating a new contribution to the energy density of the Universe as
the dark energy. Of course, more elaborated models including matter sources like dust and radiation must be studied.
\section*{Acknowledgements}
I would like to thank FACITEC (Vit\'oria-ES), CLAF/CNPq/MCT and CBPF for financial support.

\begin{thebibliography}{0}
\bibitem{acel} N. Pinto-Neto and E. Sergio Santini, Phys. Lett. {\bf A} 315:36-50 (2003).
\bibitem{SN1} Perlmutter, S. {\it et al.}, The Astronomical Journal, {\bf 116}:1009-1038 (1998); Nature (London) {\bf 391}, 51 (1998) and The Astrophysical Journal, {\bf 517}: 565-586 (1999).
\bibitem{SN} A. Riess {\it et al.}, Astron. J. {\bf 116}, 1009, (1998).
\bibitem{bohm1} David Bohm,  Phys. Rev. {\bf 85}, 166 and 180 (1952).
\bibitem{hol} P. R. Holland,
{\it The Quantum Theory of Motion: An Account of the de Broglie-Bohm
Causal Interpretation of Quantum Mechanics}
(Cambridge University Press, Cambridge, 1993).
\bibitem{fab2} R. Colistete Jr., J. C. Fabris and N. Pinto-Neto,
Phys. Rev. {\bf D62}, 83507 (2000).
\bibitem{must} N. Pinto-Neto and E. Sergio Santini, Phys.Rev. {D \bf 59}
123517 (1999).
\bibitem{cons} N. Pinto-Neto and E. Sergio Santini, Gen. Rel. and Grav. {\bf 34},
505 (2002).
\bibitem{tese} E. Sergio Santini, PhD Thesis, CBPF-Rio de Janeiro, (May 2000),
(gr-qc/0005092).
\end{thebibliography}
\end{document}